# A Logic Programming approach for Formal Verification of NetBill security and transaction protocol

**Suvansh Lal**
Dhirubhai Ambani Institute of Information and
Communication Technology.
Gandhinagar, Gujarat, India
*suvansh_lal@daiict.ac.in*

**Abstract**

*Use of formal techniques for verifying the security features of electronic commerce protocols would facilitate, the enhancement of reliability of such protocols, thereby increasing their usability. This paper projects the application of logic programming techniques for formal verification of a well referred security and transactions protocol, the NetBill. The paper uses ALSP (Action Language for Security Protocols) as an efficient formal specification language and SMODELS a model generator to formally analyze and plan attacks on the protocol.*

## 1. Introduction

Electronic commerce enables business transactions through the internet, when a direct interaction between the parties in a business deal is seemingly difficult. It operates in an open business environment where it is not mandatory for any trust relationship to hold between the engaging parties. Therefore, the protocols that facilitate the exchange of services over the internet and form the basis of a commercial transaction are inculcated with security features, which aim at minimizing the possibilities of a fraudulent activity.

Considering the security of such protocols, besides the assumption that perfect cryptographic techniques ensure inclusion of basic properties like authentication, privacy and integrity into a protocol. There are other characteristics like **atomicity**[7] that arise at a higher level and are specific to the electronic commerce protocols. Atomic transactions form the corner stone of the modern transaction processing theory[8]. Atomicity is a property which allows us to logically link multiple operations of a protocol, in such a way that either all of them are executed during a protocol run or none of them are. It is important that all electronic commerce systems ensure atomicity as most non atomic systems like Digicash[12] or First Virtua[13] are either too expensive to implement or are severely flawed[7].

Introduction of formal techniques for verification will not only ascertain the presence or absence of the above mentioned characteristics in the protocols, but would also help analysts in designing and optimizing electronic commerce systems. In view to this, the novelty of this paper is to use a logic programming[5] approach for formally specifying NetBill[9] a prominent electronic commerce protocol. Later, we analyze the claims, both *for*[11] and *against*[10] NetBill"s goods atomicity[7] and reproduce a model for an attack similar to the one proposed in [10]. The paper emphasizes on the use of ALSP (Action Language for Security Protocol)[1] and state based description of the protocol for its analysis and verification.

ALSP is an executable specification language for representing security protocols and checking violations they are vulnerable to [1][2]. It is based on logic programming, with stable model semantics [4]. Logic Programming enables one with declarative ease to specify the actions of different agents in a protocol. This includes both the operational behavior of a protocol, along with the possible security attacks of an intruder. A protocol trace is viewed as a plan to achieve a goal and the security attacks become plans, which achieve goals that correspond to security violations. A specification in ALSP, is executable using SMODELS[6] as a model generator.

A protocol trace can be written using logic programming with stable model semantics. All stable models for the solution set of logic programs are minimal and grounded in nature, where each atom has a justification in terms of the program [3]. The minimalism and grounded characteristics, make ALSP particularly suitable for specifying payment and transactions protocols; as we are able to derive exactly what happened which lead to a security violation. The model generated by executing the specification also specifies that every action, which occurred during the execution of the specification, has a justification behind it. Similarly if the execution of the specification does not reveals a model then, it can be declaratively stated that there exists no plan to achieve that violation.

Specification of a protocol in ALSP requires inculcation of concepts of robotic planning[2]. Security protocols are reframed as planning problems, where agents exchange messages and are subject to attacks by intruders. The initial state of a protocol is defined in terms of messages already exchanged by the agents and information they have derived from them. A goal state is defined as an unwanted state where an attack has already occurred. Actions are specified as exchange of messages amongst agents. If there exists a plan to achieve the goal, we state that the protocol is flawed with scope for a security violation. A solution to the planning problem if any is a sequence of steps that leads to a goal state, corresponding to an attack. A more explicit discussion on protocol specification using ALSP can be found in the works of Aiello and Massacci[3].

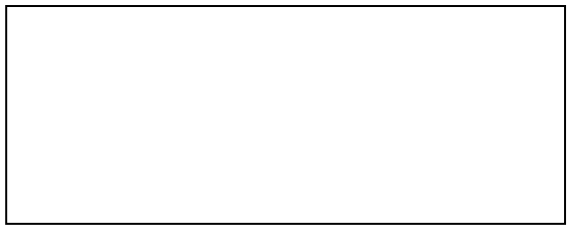

In Section 2 of the paper, we begin with a mention of some important security features for electronic commerce protocols. Section 3 deals with the explanation of the NetBill protocol and the notations used in the paper. The development of the specification for the protocol, using the logic programming techniques is explained in the Section 4. We finally specify attacks on the NetBill protocol and produce our results in Section 5, which is followed by the conclusion of our work.

## 2. Security in Electronic Commerce

Security is a primary concern for all electronic commerce transactions. The security features, as suggested by the PODC Community[14] for electronic commerce protocols include:
- atomic transactions
- cryptographically secure protocols
- secure computations
- high reliability

For our analysis of the NetBill protocol, we will focus only on the aspect of atomicity. Tygar[7] explains the importance of atomicity in electronic commerce. Atomicity as a characteristic can be further classified into three distinct levels. Each level caters to a separate security aspect in a transaction. These levels include money atomicity, goods atomicity and certified delivery [7].

### 2.1 Money Atomicity

A protocol is stated to be money atomic, if no creation or destruction of money takes place during the protocol run. In simpler terms, during a transaction in a protocol, neither of the participating agents should create nor destroy money which they intend to transfer to the other participating agent of the protocol.

### 2.2 Goods Atomicity

Goods atomic protocols effect an exact transfer of goods for money[7]. This characteristic eliminates all possibilities of an agent getting goods without paying for them or vice-versa. All goods atomic protocols are inherently money atomic.

### 2.3 Certified Delivery

Certified delivery protocols enable all engaging parties, to prove exactly what goods have been delivered through the protocol run. They are extremely helpful in scenarios where both the merchant and the customers are not trusted. All protocols that ensure a certified delivery are also goods and money atomic.

As NetBill [9] is intended to be used for the sale and purchase of information goods. Goods atomicity is an important consideration for all information goods protocols. NetBill has previously been proved to be goods atomic[11], we will test the validity of this formally using the logic programming approach and show the richness and competence of ALSP as formal verification language in verifying similar attacks on different protocols.

## 3. The NetBill Protocol

The NetBill[9] security and transactions protocol was proposed in the year 1995. NetBill is system for micropayments for information goods on the internet. It provides for strong authentication, privacy and atomicity in payment and delivery of goods. A NetBill transaction model involves three active parties namely, the customer, the merchant/vendor and the NetBill transaction server. Figure 1 gives a brief view of interaction between the different parties in a NetBill transaction model.

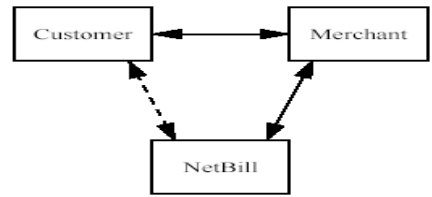

Figure 1: Parties in NetBill transaction [9]
[———] Transaction Protocol; [----] Auxiliary Protocol

### 3.1 Notations

The notations used throughout the paper to explain the NetBill protocol and its cryptographic operations are:

U, V, T are agents to the protocol.
U→V: M, can be understood as agent(U) sends the message(M) to the agent(V) in the protocol.
U: M, can be understood as agent(U) synthesizes message(M)
EG: encrypted goods
H(EG): hash value of the encrypted goods(EG)
EPO: electronic payment order
Sv(M): agent(V)''s signatures on message(M)

### 3.2 Protocol description

The protocol enables a vendor(V) and a user(U) to sell and buy information goods over the internet. The protocol being non-optimistic in nature, the NetBill transaction server(T) acts as a trusted third party and is central to the protocol run. All transactions between the user(U) and the vendor(V) take place with the help of the NetBill server(T) and it is assumed that the server(T) is always online.NetBill has three distinct phases, each of which depends on the
success of the previous phases.

The **negotiation phase** is the initial phase, which involves both the user(U) and the vendor(V) agreeing on the prices for the goods and the services to be exchanged. We will not make specific mention of these steps in our design and protocol specification as the security characteristics of privacy, authentication and atomicity are not inherent in the negotiation phase[11].

U→V: Price Quote Request
V→U: Price Quote

The second phase is a **delivery phase** which involves the user(U) sending a purchase request(for the agreed prices) to the vendor(V). The vendor(V) then encrypts the ordered goods, along with the hash of the encrypted goods and sends it back to the user(U).

U→V: Purchase Request
V: EG, H(EG)
V→U: EG, H

The **payment phase** follows the delivery phase in which the user(U) transfers a synthesized electronic payment order(EPO) to the vendor(V). The vendor(V) signs the EPO and concatenates it along with the specific key(K) with which, it had earlier encrypted the goods delivered to the respective user(U). It then forwards this information to the NetBill transaction server(T). The server(T) credits the vendor(V)"s account with the money from the EPO. It then signs the key(K) forwarded by the vendor(V) and sends it back to the vendor(V). The signature is later forwarded to the user(U) which then decrypts the goods with the relevant key(K).

U: EPO
U→V: EPO
V→T: S$_V$(EPO, K)
T→V: S$_T$(receipt, K)
V→U: S$_T$(receipt, K)

There also exists an **optional phase** where the user(U) directly communicates with the NetBill server(T), when it fails to receive the signed key from the vendor(V).

U→T: transaction enquiry
T→U: S$_T$(receipt, K)

## 4. Specification of NetBill using ALSP

ALSP is based on logic programming with stable model semantics[2]. A logic program is defined by rules. We formalize and frame all actions in a protocol specification as logic program rules[3]. Therefore every time rule holds true, an action is said to be performed and the protocol is assumed to have reached a new state. Following the logic programming approach all rules comprise of a *head* and a *body* [2]. The *head* and the *body* in each rule is separated by a [:--] to make it syntactically correct for the SMODELS model generator. The left hand side *head* literals hold true if and only if all the literals on the right hand side *body* are true. A syntactically correct example of a rule would be:

q :- not p, s.

The above rule is read as "q iff( not p && s)". Here „q", „p" and „s" are literals and we can state informally that „q" is true if and only if „p" is not true and „s" is true.

We start specifying the protocol by considering the basic sort predicates(predicates which are not governed by any rules, mentioned in the specification) first[3]. We state clearly the background theory(initial state of the protocol) which contains all the rules describing, how a message is composed and decrypted by the agents. It includes the property of keys shared and how information is attained by agents participating in a protocol. The basic sort predicates that we will use to specify the NetBill protocol are similar to ones taken from the works of Aiello and Massacci[3]. Though, a few predicates that cater specifically to the NetBill protocol are added
intentionally to its specification. They include:

```
encryptedGoodHashPair(EncGood, Hash)    (1)
goodsDescriptionQuotePair(GD, Quote)    (2)
```

While writing the specification it becomes imperative to add basic sort predicates depending on the assumptions that form the basis of the specification. For example, the definition of predicate(1) has enabled us to ensure that for every encrypted good(EncGood) there exists a hash value(Hash) which is specific to the encrypted good. Similarly predicate(2) states that for every goods description(GD) that a customer may forward to a vendor, there exists a quote(Quote) with which the vendor will reply to the customers goods request.

The list of predicates that are inculcated is immense and all of them cannot be discussed here. It is also impossible to state and explain the entire protocol specification. So, we will focus on how to write the specification based on the assumptions and the states that a protocol attains during a run. Later, a classical unit action of an agent, which includes receiving a message, deriving information from the message and then replying to it will be specified and explained using ALSP.

A protocol specification requires a description of what states the protocol reaches and what are the actions that lead to that state. To do the same, we give a phase wise state analysis of the NetBill[9] protocol. Below each state, is a set of preconditions that must hold true for that state to be attained. The notations used for protocol description are identically similar to the one described in [section 3.1].

### 4.1 The Negotiation Phase (N/P)

```
[State 1]      U→V: Price Quote Request
               Pre-Condition: None

[State 2]      V receives Price Quote Request
               Pre-Condition: [State 1]

[State 3]      V, synthesizes a Price Quote
               Pre-Condition: [State 2]

[State 4]      V→U: Quote
               Pre-Condition: [State 2, 3]

[State 5]      U, receives Quote from V
               Pre-Condition:[State 4]

[State 6]      U, accepts transaction with V
               Pre-Condition:[State 5]
```

Now, for the sake of more clarity, we will see how an action is specified in ALSP. Let us consider the States(2), (3) and (4) of the negotiation phase. These states together comprise of a unit action of a vendor(V) which includes receiving a price quote request, synthesizing a quote and sending the quote to the relevant customer(U). The nomenclature of all the literals and atomic functions in the specification is highly intuitive and is indicative of the action an agent takes when the protocol is in the desired state. The specification of the states(2), (3) and (4) in ALSP would be:

```
receivedQuoterequest(V,QuoteRequest):-
     sendQuoterequest(U, V, QuoteRequest),
     vendor(V),user(U),
     quotereq(QuoteRequest).           (A)
```

```
synthesizePriceQuote(V, Price):-
      receiveQuoterequest(V, QuoteRequest),
      vendor(V), priceValue(Price),
      quotereq(QuoteRequest).         (B)

sendQuote(V, U, Price):-
      synthesizePriceQuote(V, Price),
      receivedQuoterequest(V,QuoteRequest),
      user(U), vendor(V),
      priceValue(Price).              (C)
```

      The rule(A) ensures that a state in which the vendor(V) receives a quote request is true if and only if, the user(U) had earlier send the quote request to the vendor(V). Which would be true, only if there existed a valid user(U), a vendor(V) and a quote request.

      The rule(B) ensure that the state in which a vendor(V) synthesizes a price quote is true if and only if, the vendor(V) has earlier received a quote request from the user(U).

      The rule(C) ensures that the state in which a vendor(V) sends a price quote to the user(U) is true if and only if, he has synthesized a price quote for the respective user(U) from which he had received a quote request. Similarly all actions for the Negotiation Phase can be specified easily in ALSP.

### 4.2  The Delivery Phase (D/P)

```
[State 1]     U→V: Purchase Request
              Pre-Condition: [State 6(N/P)]

[State 2]     V, receives the purchase request
              Pre-Condition: [State 1]

[State 3]     V, stores the purchase request
              Pre-Condition: [State 2]

[State 4]     V, generates a random key
              Pre-Condition: [State 3]

[State 5]     V, encrypts the goods which it
              intends to send corresponding to
              the purchase request.
              Pre-Condition: [State 3, 4]

[State 6]     V, generates hash corresponding
              to the encrypted goods
              Pre-Condition: [State 5]

[State 7]     V→U: Sends encrypted goods and
              hash of goods
              Pre-Condition: [State 5, 6]

[State 8]     U, receives the encrypted goods
              and the hash value
              Pre-Condition: [State 7]

[State 9]     U, compares hash value with the
              encrypted goods
              Pre-Condition: [State 8]
```

      The send and receives states for the user(U) and the vendor(V) are similar in nature to that of the Negotiation Phase. To enable encryption by the vendor(V) we have to specify separately his ability to generate a random key and store it for further transactions. It can be inherently assumed in the protocol specification that a valid hash function has been pre-established between the user(U) and the vendor(V) and is outside the scope of the protocol.

### 4.3  The Payment Phase (P/P)

```
[State 1]     U, generates the Electronic
              Payment Order(EPO)
              Pre-Condition: [State 6(N/P),
              9(D/P)]

[State 2]     U→V: EPO
              Pre-Condition: [State 1]

[State 3]     V, receives EPO from U
              Pre-Condition: [State 3]

[State 4]     V, signs EPO + Key
              Pre-Condition: [State 3, 2(D/P)]

[State 5]     V→T:  Sv(EPO,  Key)
              Pre-Condition: [State 4]

[State 6]     T, receives Sv(EPO,  Key)
              Pre-Condition: [State 5]

[State 7]     T, signs the Key with his
              signatures
              Pre-Condition: [State 6]

[State 8]     T→V:  ST(receipt, Key)
              Pre-Condition: [State 6, 7]

[State 9]     V,receives the signed Key from T
              Pre-Condition: [State 8]

[State 10]    V→U:  ST(receipt, Key)
              Pre-Condition: [State 9]
```

      The generation of the Electronic Payment Order(EPO) by the user(U) depends on the information saved by user at the time of accepting the transaction in the Negotiation Phase(N/P). The EPO is generated only after a successful comparison of the hash value with the encrypted goods transmitted during the Delivery Phase(D/P). The ALSP specification for the generation of the EPO would be:

```
electronicPaymentOrder(sk(U), GoodDescription,
Price, hash(f, encryptGoods(K, Goods))):-
      comparesHash(U),
      userKnowsEncryptedGoods(U, encGoods(K,
      Goods)),
      userKnowsHash(U,hash(f,encryptGoods( K,
      Goods))).
                                          (D)
```

      The rule(D) ensures that the synthesis of the Electronic Payment Order occurs if and only if the user(U) has a knowledge of the encrypted goods; the hash value transmitted by the vendor(V) and the user(U) has successfully compared the two together. It is evident from the *head* of the rule that the EPO contains a good description, the money to be debited and the hash value of the encrypted goods which is later forwarded to the NetBill Transaction server(T).

### 4.4  The Optional Phase (O/P)

```
[State 1]     U→T: transaction enquiry
              Pre-Condition:[State 10(P/P)]
```

```
[State 2]      T→U: ST(receipt, Key)
               Pre-Condition:[State 1]
```

The first step of the Optional Phase(O/P) occurs when the user(U) is unable to get the decrypting key(K) from the vendor(V). The user(U) then enquires the NetBill server(T) about the transaction. The NetBill server(T) in response to the transaction enquiry resends the signed key(K) and the receipt of the transaction with the vendor(V), directly to the user(U). The formal specification of the optional transaction enquiry in ALSP would be:

```
transactionEnquiry(U, T, transactionenquiry,
Time):-
     not receiveUserReceipt(U,receipt (
     sk(T), K, computedReceipt(R)), Time)
                                              (E)
```

The rule(E) ensures that a transaction enquiry from user(U) to the NetBill server(T) is true if and only if the user(U) has not received the signed receipt and the key(K) from the server(T) in time. This completes the specification of the protocol.

But, a model for a security violation also needs a goal state that corresponds to that attack[3]. While planning an attack on a protocol, we must concentrate on a security guarantee(like authentication, privacy, atomicity), a protocol intends to offer, which we aim to prove is void. An ALSP specification that challenges that guarantee is tested as a goal state. If there exists a model for the challenge, we declaratively conclude that protocol is flawed and there is scope for a violation which corresponds to our specified goal state.

## 4 . Specification of Attacks on NetBill

NetBill protocol is used for purchase and sale of information goods[9]. It is imperative for an information goods protocol to be goods atomic[7]. There have been claims both for and against the goods atomicity of the NetBill protocol. Researchers Ogata and Futasugi[10] claimed that the NetBill is not goods atomic. While specifying the goal state, we are following their assumptions about the protocol and check whether our approach is competent enough to
specify formally the attack proposed by them.

In order to define a goal state that would correspond to security violations challenging the goods atomicity of the protocol, we would have to enable the user(U) to judge a valid decryption of the goods he ordered at the end of the protocol run. To do the same, we add an extra rule in our specification which would judge the validity of the decrypted goods.

```
userSuccessfullDecryption(U, Goods):-
     K1=K,
     knowsReceipt(U, sk(T), K1,
     computedReceipt(R)),
     userKnowsEncryptedGoods(U,
     encryptGoods(K, Goods)).          (F)
```

The rule(F) of the protocol specification ensures that there exists a successful decryption of the goods(Goods) if and only if the key(K1) received from the server(T) is the same key(K) with which the vendor(V) had encrypted the goods which were transferred to the user(U).

An attack claiming that NetBill is not goods atomic would define a scenario in which the user(U) pays for the goods, but is unable to receive correct goods which it had ordered prior to the payment. Using the concepts of logic programming and planning, we finally specify our goal state.

```
Attack(U) :-
     not userSuccessfullDecryption(U, Goods,
     knowsReceipt(U, sk(T), K1,
     computedReceipt(R)),
     electronicPaymentOrder(sk(U),
     GoodDescription, Price, hash(f,
     encryptGoods(K, Goods))).
```

The "Attack(U)" of the specification explains a state in which the user is unable to successfully decrypt the goods at the end protocol run. Although for the same transaction the user(U) had generated an EPO for which a corresponding endorsed receipt(signifying a debit) has been received from the server(T).

We now use the SMODELS[6], model generator to test the specification and see if there exists an attack on NetBill. The results vary based on our assumptions about the honesty of the vendor(V) in the protocol. Therefore they have been categorized into the following cases.

CASE 1: The vendor is assumed to be honest and does not transmit two different keys for the same transaction.

```
MODEL GENERATED:        0
userSuccessfullDecryption(U, Goods)  :   TRUE
Attack(U)                            :   FALSE
```

CASE 2: The vendor is assumed to be dishonest and is capable of transmitting two different keys for the same transaction.

```
MODEL GENERATED:        1
userSuccessfullDecryption(U, Goods)  :   FALSE
Attack(U)                            :   TRUE
```

We are able get a model for our assumption in Case 2. This result corresponds to the result proposed by Ogata and Futasugi[10]. A similar approach for specification can be adopted for other security and transactions protocols as well. Although our results show that NetBill is not goods atomic, it cannot be stated declaratively, as NetBill has a devised contingency mechanism for similar security threats which is external to the protocol scope.

## 5. Conclusion

The logic programming approach for formal verification of security protocols, unlike other methodologies prevents analysts from learning scripting styles that are specific to those methodologies. The syntax is intuitive and generic in nature, which makes specifications more comprehensible to people. Besides, the approach can be used in formal verification of a wide plethora of protocols. A problem with this approach may be that the specification for complex protocols becomes very lengthy and time taking. But, being executable in nature, it enables us to study a generated model, based on the specification of an attack and determine exactly what went wrong during the protocol run. This property comes handy in not only determining security threats, but also designing and optimizing protocols.

Suvansh Lal. DA-IICT. Gandhinagar. Gujarat.